\documentclass[prl,twocolumn,amsmath,amssymb,superscriptaddress]{revtex4-1}
\usepackage{graphicx}
\usepackage{natbib,hyperref}
\usepackage{bm}

\usepackage{nccmath}

\usepackage[normalem]{ulem}
\usepackage{color}

\begin{document}

\title{Interplay between friction and spin-orbit coupling as a source of spin polarization}

\author{Artem G. Volosniev}
\affiliation{IST Austria (Institute of Science and Technology Austria), Am Campus 1, 3400 Klosterneuburg, Austria}

\author{Hen Alpern} 
\affiliation{Applied Physics Department and the Center for Nanoscience and Nanotechnology, The Hebrew University of Jerusalem, 91904 Jerusalem, Israel}
\affiliation{Racah Institute of Physics and the Center for Nanoscience and Nanotechnology, The Hebrew University of Jerusalem, 91904 Jerusalem, Israel}

\author{Yossi Paltiel} 
\affiliation{Applied Physics Department and the Center for Nanoscience and Nanotechnology, The Hebrew University of Jerusalem, 91904 Jerusalem, Israel}

\author{Oded Millo} 
\affiliation{Racah Institute of Physics and the Center for Nanoscience and Nanotechnology, The Hebrew University of Jerusalem, 91904 Jerusalem, Israel}

\author{Mikhail Lemeshko}
\affiliation{IST Austria (Institute of Science and Technology Austria), Am Campus 1, 3400 Klosterneuburg, Austria}

\author{Areg Ghazaryan}
\affiliation{IST Austria (Institute of Science and Technology Austria), Am Campus 1, 3400 Klosterneuburg, Austria}

\begin{abstract}
We study an effective one-dimensional quantum model that includes friction and spin-orbit coupling (SOC), and show that the model exhibits spin polarization when both terms are finite. Most important, strong spin polarization can be observed even for moderate SOC, provided that friction is strong. Our findings might help to explain the pronounced effect of chirality on spin distribution and transport in chiral molecules. In particular, our model implies static magnetic properties of a chiral molecule,
which lead to Shiba-like states when a molecule is placed on a superconductor, in accordance with recent experimental data.
\end{abstract}

\maketitle
The second law of thermodynamics is not symmetric with respect to the time reversal, and implies irreversibility of thermodynamic processes even if the underlying fundamental laws of nature enjoy time reversal symmetry. Therefore, when considering an open system, it is natural to question the stability of features protected by microscopic time-reversal symmetry. It was recently argued that interactions between a spin and its environment may lead to time-reversal-breaking processes, which are forbidden in isolated systems~\cite{McGinley2020}, and that even the degeneracy due to the Kramer's theorem can be lifted~\cite{Lieu2021}.
This shows that properties of an open system with spin may be very different from those of a closed system. 

In this paper, we study breaking of the Kramer's degeneracy theorem, which originates from the interplay between SOC and a basic dissipative process -- frictional dissipation -- in a one-dimensional setting.
We illustrate our general findings
by discussing the spin sensitivity of chiral molecules -- a phenomenon observed in many experiments, which lacks a satisfying theoretical explanation.
A number of theoretical models have been developed~\cite{Yeganeh2009,Medina2012,Varela2013,Guo2012,Gutierrez2012,Gutierrez2013,Guo2014,Ortix2015,Matityahu2016,Michaeli2019,Yang2019, Geyer2019,Gersten2013,Dalum2019,Fransson2019, Michaeli2019, Ghazaryan2020,Du2020,Ghazaryan2020filtering, Utsumi2020, Fransson2021,Liu2021} to understand this sensitivity in connection to 
the chiral induced spin selectivity (CISS) effect~\cite{Gohler2011, Naaman2015, naaman2019_review}.
However, certain key questions remain unanswered. In particular, the CISS effect is too pronounced to be caused by weak SOC of light atoms that constitute organic molecules. In our work, SOC and friction act in unison, so that weak SOC can effectively be enhanced by dissipation.

\begin{figure}
\includegraphics[width=8.5cm]{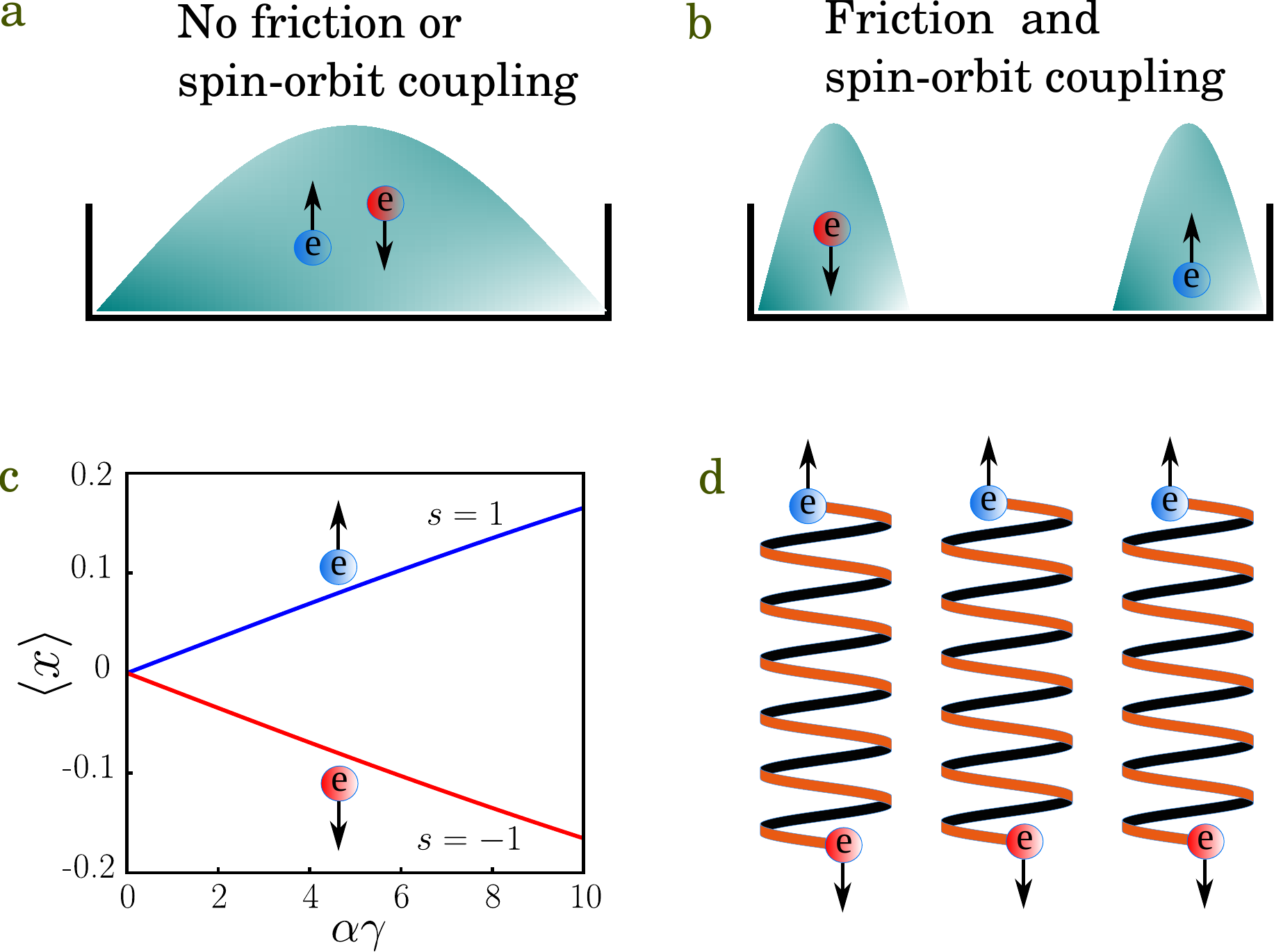}
\caption{(a) and (b) An illustration of the effective model introduced in Eq.~(\ref{HamGen}). (a) Without friction ($\gamma=0$) or spin-orbit coupling ($\alpha=0$), the spatial distribution of spins in a quantum wire is mirror symmetric. (b) If both terms are included, spin-up and spin-down particles move in opposite directions. (c) The average position of spins in the Albrecht's model of dissipation, as a function of the only parameter in the model, $\gamma\alpha$. Here, we use dimensionless quantities defined in the text. (d)~For chiral molecules on a substrate, this separation might imply magnetic properties.}
\label{fig:figure1}
\end{figure}

In the absence of dissipation, the standard route to study the motion of a particle in a medium is to map the problem onto some effective single-particle Hamiltonian. This approach is at the heart of our understanding of electrons in materials~\cite{Kittel1953}. There is no simple strategy of adding dissipation to such effective models, since a naive quantization of classical dissipative terms is incompatible with the Heisenberg uncertainty relation, as can be demonstrated using a simple harmonic-oscillator model~\cite{Senitzky1960}. It is possible to include a quantum analogue of the classical frictional force in time evolution of the one-body density matrix~\footnote{We focus on friction due to the resistance of the viscous medium. The non-contact friction~\cite{Persson2007} is not relevant for the present work. }. The corresponding term can either be derived using solvable models (e.g., the Caldeira-Leggett model~\cite{Caldeira1983}) or matched to the classical Stokes' drag, $-\gamma p$. A somewhat simpler approach is to work with dissipative Schr{\"o}dinger equations~\cite{Kanai1948,Kostin1972,Hasse1975,Albrecht1975}, which are constructed from the classical limit and satisfy certain conditions, e.g., the Heisenberg uncertainty principle, and the Ehrenfest theorem. This way is purely phenomenological and leads to a number of possible modified Schr{\"o}dinger equations. However, they are compatible with each other and with master-equation approaches, at least for simple systems~\cite{Hasse1975,Caldirola1982,Sun1995,Schuch1999}.

In this work, we employ a modified Schr{\"o}dinger equation to study the interplay between frictional dissipation (that emerges from electron-bath interactions) and SOC.
We assume that the timescale of the spin coherence is very large, i.e., there are no spin-changing interactions in the system (``spin flips'').  This assumption allows us to use the eigenstates of the modified Schr{\"o}dinger equation to understand the spin distribution of an electron confined in a well. Our main finding is that the spin-momentum locking leads to a position-dependent spin distribution: Spin-up particles are spatially separated from spin-down particles. We speculate that the predicted effect has been observed with chiral molecules adsorbed on various substrates.  For example, it has been shown that chiral molecules adsorbed onto $s$-wave superconductors induce triplet superconductivity~\cite{Alpern2016,Shapira2018,Alpern2019}, or Shiba-like states~\cite{Alpern2019}; chiral molecules on ferromagnetic surfaces induce magnetic ordering~\cite{BenDor2017,Sukenik2020}.

\textit{Theoretical model.}  Our work rests on the two observations: (i) A spin-orbit interaction couples particle's spin to its momentum; (ii) the classical frictional force is determined by the direction of the momentum. The observations suggest that frictional dissipation could potentially couple to spin, hence, lead to spin currents, and steady states that do not respect the Kramers degeneracy theorem. To investigate these possibilities, we study a quantum wire, i.e., a one-dimensional (1D) finite system, with spin-orbit coupling, $\alpha p\sigma_z$, where $\alpha$ is the SOC amplitude, $p$ is the momentum operator, and $\sigma_z$ is the Pauli matrix. Without loss of generality, we choose SOC in the $\hat{z}$-direction.  The 1D SOC interaction does not couple spin-up and spin-down particles; we define the corresponding quantum number as $s=\pm 1$. To include friction for a given $s$, we use  a phenomenological potential, $\gamma W$, where $\gamma>0$ and $W$ determine, respectively, the strength and the spatial distribution of frictional dissipation~\cite{Hasse1975}. $W$ depends on the state of the system, so that the resulting time evolution is non-linear.
Motion of a particle of mass $m$, and spin $s$ obeys the following equation
\begin{equation}
i\hbar \frac{\partial\Psi_s}{\partial t}=H_s\Psi_s; \quad H_s=\frac{p^2}{2m}+\alpha p s+V\left(x\right)+\gamma W,
\label{HamGen}
\end{equation}
where we assume a box potential: $V(x)=0$ for $|x| \leq a$, and $V\to\infty$ otherwise. A clear advantage of Eq.~(\ref{HamGen}) over other models of dissipation is its simplicity. It will allow us to analyze the problem analytically.  

If $\gamma=0$, then the eigenstates of Eq.~(\ref{HamGen}) are $\Psi_s=e^{-i\left(E_0t+m\alpha s x\right)/\hbar}\sin(\pi n(x+a)/(2a))$, where $n$ is integer and $E_0=\hbar^2\pi^2n^2/8ma^2+m\alpha^2/2$; $|\Psi_s|^2$ is illustrated in Fig.~\ref{fig:figure1}~(a) for $n=1$. For each value of~$s$, there is a spin current, which, however, does not lead to any observable flux when we average over $s$ (cf.~Ref.~\cite{Rashba2003,Fajardo2017}). If $\gamma\neq 0$ and $\alpha=0$, $s$ does not enter Eq.~(\ref{HamGen}), and there can be no interesting effects associated with spin. 

Let us now consider a system with non-vanishing SOC and friction. We focus on the Albrecht's potential~\cite{Albrecht1975},  
\begin{equation}
W=\langle p\rangle_s\left(x-\langle x\rangle_s\right),
\end{equation}
where $\langle O\rangle_s=\int\Psi_s^* O\Psi_s\mathrm{d}x$.  Time evolution of quantum averages that obey Eq.~(\ref{HamGen}) is connected to classical time evolution:
\begin{equation}
\left\langle \frac{d H_s}{d t}\right\rangle_s = -\gamma\langle p \rangle_s\frac{d\langle x \rangle_s}{dt}, \qquad \frac{d\langle x \rangle_s}{dt}=\frac{\langle p \rangle_s}{m}+\alpha s,
\label{eq:classical_limit}
\end{equation}
which explicitly demonstrates that the potential $W$ leads to energy dissipation, and breaking of the time-reversal symmetry. Solutions of Eqs.~(\ref{eq:classical_limit}) are time-independent if $\langle p \rangle_s=-\alpha ms$~\footnote{Note that the average value of $p$ is zero for $\alpha=0$. In this case, there is no frictional force, i.e., $W=0$, and the steady-state solutions of $H$ are simply the eigenstates of a square well. Therefore, non-vanishing spin-orbit coupling is essential for the main results of present work.}. 
The system is dissipative, therefore, only the `lowest-energy' steady-state of $H_s$ can be interpreted as the fixed point of time evolution, see, e.g., Ref.~\cite{Albrecht1975}. Below, we explore the physical nature of this state. In what follows, we use dimensionless units: $\tilde{x}=x/a$, $\tilde{t}=\hbar^2 t/(2ma^2)$, $\tilde{\alpha}=2m\alpha a/\hbar$ and $\tilde{\gamma}=2m\gamma a^2/\hbar$. For simplicity, we shall omit tilde.

To find the steady state of the Albrecht's model with SOC, we apply the gauge transformation: $\Psi_s(x,t)=e^{-\frac{i\alpha s x}{2}-i E t}f_s(x)$, which transfers the SOC term into renormalization of dissipation. We use the steady-state value of $\langle p \rangle_s$ to derive the equation for $f_s$
\begin{equation}
-\frac{\partial^2 f_s}{\partial x^2}+V(x)f_s+\beta xf_s=E'f_s,
\label{Eq:f_s}
\end{equation}
where $E'=E-\alpha^2/4+\beta \langle x\rangle_s$. Note that the defining parameter here is $\beta=-\gamma\alpha s/2$, which means that the effect of the linear potential in Eq.~(\ref{Eq:f_s}) can be large even for weak SOC. In this sense, SOC and friction are interchangeable in this model: the steady state obtained with weak friction and strong SOC is identical to that with strong friction and weak SOC. This happens because the frictional force is determined by $\gamma \langle p\rangle_s$, where the value of $\langle p\rangle_s$ is dictated by the SOC strength, $\alpha$.

Equation~(\ref{Eq:f_s}) is the Airy (Stokes) equation whose solution is
\begin{align}
f_s=\kappa_1 \mathrm{Ai}\left[\sqrt[3]{\beta}\left(x-x^0_s\right)\right]+
\kappa_2 \mathrm{Bi}\left[\sqrt[3]{\beta}\left(x-x_s^0\right)\right],
\label{eq:fx}
\end{align}
where $x_s^0=E'/\beta$, $\mathrm{Ai}$ and $\mathrm{Bi}$ are Airy functions of the first kind and the second kind, respectively. The parameters $\kappa_1$,$\kappa_2$ and $E^\prime$ are determined from the boundary conditions due to $V$ and normalization condition; $\langle x\rangle_s$ and $E$ that enter $E'$ are determined self-consistently. 

The Hamiltonian~(\ref{HamGen}) commutes with $U_0P$, where $U_0=e^{i\sigma_x\pi/2}$ is the spin rotation and $P$ is the inversion operator. Therefore, the eigenstates for $s=\pm 1$ are degenerate so that $\langle x\rangle_{+1}=-\langle x\rangle_{-1} $\footnote{If the potential $V$ is not mirror symmetric, then there is no double degeneracy associated with spin. This occurs due to dissipation, which explicitly breaks time-reversal system, and brings the system outside the region of applicability of the Kramers degeneracy theorem.}. The sign of $\langle x\rangle_s$ is determined by the sign of $\alpha s$. Figure~\ref{fig:figure1}~(c) shows the mean value of the position operator, $\langle x\rangle_s$, as a function of  $\gamma\alpha$. Opposite spin states are separated in space, and this separation increases linearly for small values of~$\gamma\alpha$. The property $\langle x\rangle_{+1}=-\langle x\rangle_{-1}$ is also clearly observed. Note that $\langle x\rangle$ is measured in units of $a$, which means that the absolute strength of the effect is larger for longer systems provided that the spin coherence length is larger than $a$.     

If the system is initialized in some parity-symmetric state, then at $t>0$ the frictional force will drive particles with different spins into opposite directions, leading to an assymetric state.  
Figure~\ref{fig:figure1}~(b) illustrates the solution to the static problem of Eq.~(\ref{eq:fx}), which we expect to define a fixed point of time evolution. The amplitude of the frictional force is given by $\gamma\alpha$, and therefore, a strong dissipative system can act as an efficient spin filter even for weak SOC. This is not an artifact of the specific (Albrecht's) form of $W$. We have observed similar physics for other standard (e.g., S\"ussmann's and Kostin's) forms of $W$~\footnote{See Ref.~\cite{Hasse1975} for a detailed comparison of different phenomenological models of friction for systems without SOC.}, and for traditional master-equation approaches to friction. A detailed analysis of different models with SOC and friction will be published elsewhere.  

In this work, we do not attempt to derive from the first principles the effective model introduced in Eq.~(\ref{HamGen}). However, we must outline a few necessary conditions for the validity of that model. First of all, it {\it should not} be possible to gauge out the SOC term from the microscopic Hamiltonian that underlies Eq.~(\ref{HamGen}). In one spatial dimension, this implies that  beyond-1D effects must be included, see, e.g., Ref.~\cite{Moroz1999}. The microscopic Hamiltonian may in principle allow for ``spin flips'' in two-body scattering, however, their rate should be small in comparison to spin-preserving collisions, i.e., the spin coherence length must be much larger than the mean-free path of a particle.  This requirement is needed as our effective model preserves spin. 

\textit{Spin separation in chiral molecules.} We now consider a helical molecule, in which the kinetic energy of an electron is transformed (dissipated) into a vibrational motion of atoms that constitute the molecule and the substrate. Note that spin and charge reorganization have already been observed in a layer of chiral molecules by using a modified Hall device~\cite{Kumar2017} and, therefore, a chiral molecule is a candidate for observing the physics described above.

We start with the 1D tight-binding model of a helical molecule with $M$ sites~\cite{Matityahu2016}
\begin{align}
H_\mathrm{mol}=\epsilon_0\sum_{m=1}^{M}c^\dagger_{m}c_{m}-
J\sum_{m=1}^{M}\left[c^\dagger_{m+1}V_mc_{m}+\mathrm{H.c.}\right],
\label{HamHelix}
\end{align}   
where $c^\dagger_m=(c^\dagger_{m,\uparrow},c^\dagger_{m,\downarrow})$ is the creation operator at the site $m$, $V_m=e^{i\mathbf{K}_m\cdot\boldsymbol{\sigma}}$ is a unitary matrix which defines SOC, and $J$ is a hopping amplitude. The parameter $\mathbf{K}_m$ reads as
\begin{equation}
\mathbf{K}_m=\frac{\lambda}{l}\left[\frac{h}{N}\left(-S_m\hat{\mathbf{x}}+C_m\hat{\mathbf{y}}\right)-2R\sin\left(0.5\Delta\varphi\right)\hat{\mathbf{z}}\right],
\label{SOCK}
\end{equation}
where $l=\sqrt{\left(h/N\right)^2+\left[2R\sin\left(0.5\Delta\varphi\right)\right]^2}$ is the distance between nearest neighbors in a helical molecule of radius $R$ and pitch $h$; $S_m=\sin\left[\left(m+0.5\right)\Delta\varphi\right]$, $C_m=\cos\left[\left(m+0.5\right)\Delta\varphi\right]$, where $\Delta\varphi=\pm2\pi/N$ is the twist angle between nearest neighbors (the notation $+$ ($-$) corresponds to a right (left) helix, respectively), $N$ is the number of sites in each turn, and $\lambda$ is a dimensionless quantity that parametrizes SOC on a lattice. It is clear that $V_m$ is periodic after each turn of the helix $V_N=V_0$.

To connect $H_\mathrm{mol}$ to $H_s$ with $\gamma=0$, we perform the gauge transformation $c^\dagger_m=a^\dagger_m {\cal V}^{m^\prime/N}V_NV_{N-1}\dots V_{m^\prime}$, with $m^\prime=m\,\mathrm{mod}\,N$ and ${\cal V}=V_NV_{N-1}\dots V_1$, which brings the Hamiltonian~(\ref{HamHelix}) into a translationally invariant form. 
We impose periodic boundary conditions, and write the Fourier transform of the Hamiltonian as:
\begin{equation}
H(k)=\epsilon_0-2J\cos\left(k-\frac{\theta\hat{\mathbf{n}}\cdot\boldsymbol{\sigma}}{N}\right),
\label{HamHelixk}
\end{equation}
where $k=2\pi r/M$ $\left[r=0,\dots,M-1\right]$ is the wave vector in units $l^{-1}$.
The operator ${\cal V}$ now reads as ${\cal V}=e^{i\theta\hat{\mathbf{n}}\cdot\boldsymbol{\sigma}}$, where $\hat{\mathbf{n}}$ determines the quantization direction for particle's spin. 
Taking into account that SOC is expected to be weak for organic molecules ($\theta/N\ll 1$), and considering the infrared limit ($k\to 0$),  we cast the Hamiltonian~(\ref{HamHelixk}) into the form~(\ref{HamGen}) with $m=\hbar^2/(2Jl^2)$ and $\alpha=\theta \hbar/(N m l)$, where the phenomenological Albrecht's potential accounts for dissipation in the molecule.
According to our effective model, the edge of the molecule becomes spin polarized in the $\mathbf{n}$-direction, see Fig.~\ref{fig:figure1}~(d). In other words, the charge distribution of donor electrons in a chiral molecule is accompanied by spin polarization in the $\mathbf{n}$-direction determined by the sign of SOC. For weak SOC ($|\mathbf{K}_m|\ll1$), we can work in the linear regime [i.e., $e^{X}\simeq 1 + X$], which leads to $\hat{\mathbf{n}}||\hat{\mathbf{z}}$ and $\theta/N\simeq -\left(2R\lambda/l\right)\sin\left(0.5\Delta\varphi\right)$. This directly shows that SOC will have opposite values for right-handed and left-handed enantiomers, thereby facilitating spin-polarization in opposite directions.

\begin{figure}
\includegraphics[width=8.5cm]{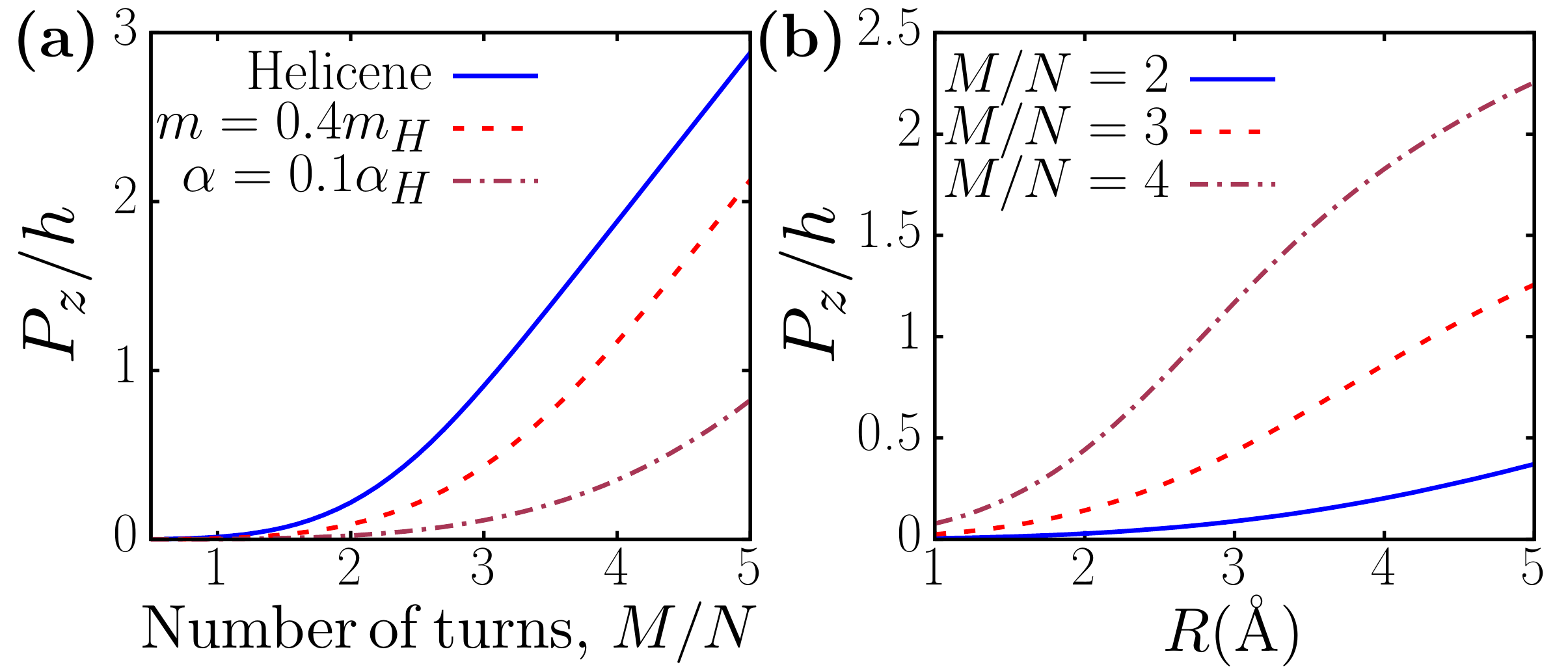}
\caption{\label{fig:figure2} Spin polarization as a function of (a) the number of turns, $M/N$, and (b) the radius, $R$, of the helix. Not-specified parameters are taken from our model of helicenes, see the text.
}\end{figure}

In our model, SOC and dissipation effectively turn a chiral molecule into a magnet, even in equilibrium. To illustrate the effect for realistic parameters, we consider a model of a helicene molecule. It was noted that helicene can be modeled as a helix with $R=3\mathrm{\AA}$, and $h=3.5\mathrm{\AA}$~\cite{Kettner2018}. The hopping amplitude $J\lambda$ can be estimated to be around 1 meV~\cite{Kettner2018}. Therefore, the SOC amplitude is $\hbar\alpha_H=4JR\lambda\sin\left(0.5\Delta\varphi\right)=5.2\,\mathrm{meV\cdot\AA}$ for $N=7$ heptahelicene. To estimate $\gamma$, we use classical linear response theory: $\gamma=e M/\mu$~\cite{Nitzan2006Book}, where $M$ is the mass of a charge carrier, and $\mu$ is the mobility.  The highest hole mobility measured in thiahelicene is $\mu\approx 2.1\,\mathrm{cm^2V^{-1}s^{-1}}$~\cite{Zhou2010,Shen2012}. The effective mass of the holes in a helicene-based semiconducting glassy film was computed to be in the range $m_H\simeq 5 m_e$~\cite{Xu2019}, where $m_e$ is the electron mass. Although these systems differ from each other, they allow us to estimate the strength of frictional dissipation as $\gamma\approx 4.19\times10^{14}\,\mathrm{s}^{-1}$. 
Using this estimate, we calculate spin polarization per pitch in the direction of the molecular axis (defined as $P_z/h=\left(\langle x\rangle_{+1}-\langle x\rangle_{-1}\right)/Nl$), see Fig.~\ref*{fig:figure2}.  The figure demonstrates that our model leads to strong polarization for helicene, even with the reduced SOC amplitude $\alpha=0.1\alpha_H$ or the effective mass $m=0.4m_H$ (which also reduces the SOC amplitude). Increasing the number of turns of the molecule results in larger polarization, which is consistent with experimental data. Figure~\ref{fig:figure2}~(b) shows that spin polarization is an increasing function of the radius of the helix 
for fixed $h,M,N$.

Below, we discuss Shiba-like states~\cite{Shiba1968} observed in superconductors coated with chiral molecules whose density is low~\cite{Alpern2019}. We shall argue that our model provides a possible explanation of the experiment.  Note that short chiral molecules (cysteine)
did not induce any sub-gap features in the experiment of Ref.~\cite{Shapira2018}, which is consistent with our prediction that the strength of spin polarization is controlled by the length of a system.

\textit{Chiral molecules on a superconductor.} We are interested here in properties of a two-dimensional (2D) superconductor coupled to a chiral molecule. To make the discussion in this section as simple as possible, we introduce the two-site model of a molecule
\begin{equation}
H_2=-J(c^\dagger_{2}v c_{1}+c_1^\dagger v^\dagger c_2)
+\gamma\alpha\sigma_z(c^\dagger_{2}c_{2}-c^\dagger_{1}c_{1}),
\label{HamTwoSite}
\end{equation}
with $v=e^{i\alpha\sigma_z}$. The first term on the right-hand-side of Eq.~(\ref{HamTwoSite}) is taken from Eq.~(\ref{HamHelix}); for simplicity, we choose $\epsilon_0=0$. The second term is taken from the lattice representation of Eq.~(\ref{Eq:f_s}). The coupling $v$ is not important as we can choose a gauge transformation to eliminate it. Non-trivial effects of SOC are contained in the last term of Eq.~(\ref{HamTwoSite}) whose sign is determined by chirality.  

To describe the superconductor, we consider the 2D tight-binding model (see, e.g., Ref.~\cite{Zhu2016book})
\begin{align}
	H_{\mathrm{sup}}=&\sum_{i,j=1}^{N_s}\left[\epsilon_S\tau_z+\Delta\tau_x\right]\sigma_0b^\dagger_{i,j}b_{i,j}\nonumber \\
	&-J_S\sum_{\langle i,j,i^\prime,j^\prime\rangle=1}^{N_s}\left[\tau_z\sigma_0b^\dagger_{i,j}b_{i^\prime,j^\prime}+\mathrm{H.c.}\right],
\end{align}
where $b^\dagger_{i,j}=(b^\dagger_{i,j,\uparrow},b^\dagger_{i,j,\downarrow},-b_{i,j,\downarrow}, b_{i,j,\uparrow})$ is the creation operator at the $(i,j)$th site of the superconductor, $b^\dagger_{i,j,\uparrow}$ is the bare operator, $\tau$ acts in a particle-hole sector, $\epsilon_S$ is the on-site energy, $J_S$ is the hoping amplitude, and $\Delta$ is the $s-$wave superconducting gap. Finally, the central site of the superconductor is coupled to the molecule via 
\begin{equation}
	H_{\mathrm{int}}=-J_I\left[b^\dagger_{\frac{N_S+1}{2},\frac{N_S+1}{2}}c_1+c^\dagger_1b_{\frac{N_S+1}{2},\frac{N_S+1}{2}}\right],
\end{equation}
where $J_I$ is the hopping amplitude between the superconductor and the molecule, and,  for convenience, we assumed that $N_S$ is odd.

\begin{figure}
\includegraphics[width=8.5cm]{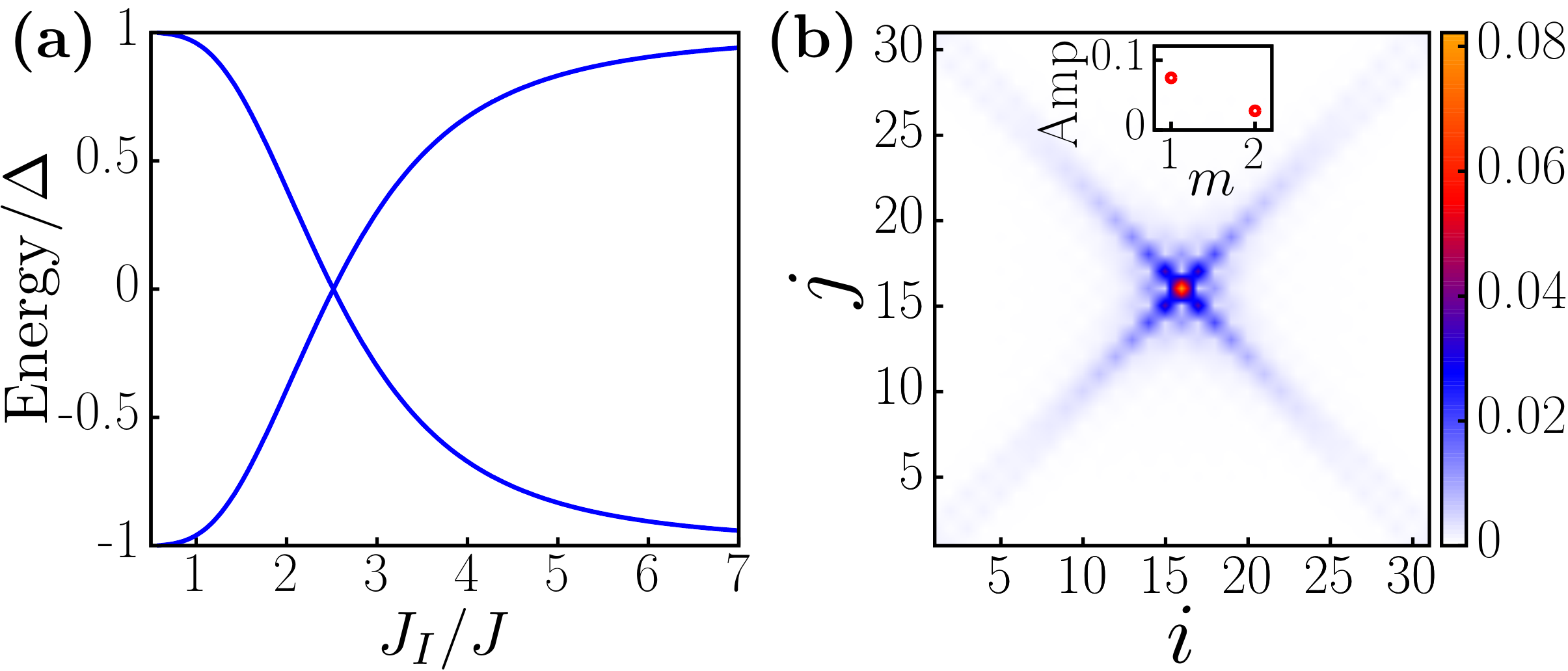}
\caption{\label{fig:ShibaTwoSites} (a) Energy levels inside the superconducting gap for the two-site molecule~(\ref{HamTwoSite}) on top of a superconductor as function of the hopping amplitude, $J_I$. (b) The probability amplitude of the lowest Shiba-like states in the superconductor (main) and in the molecule (inset) when $J_I/J=2.0$. Other parameters are $J_S/J=2.0$ and $\Delta/J=0.2$ (we measure the parameters of the Hamiltonian in units of $J$), $\epsilon_S=0.0$, $\alpha\gamma=\pi/2$. The superconductor is comprised of $31\times 31$ lattice sites, i.e., $i\in [1,31]$ and $j\in [1,31]$. We employ periodic boundary conditions.
}\end{figure}

We solve $H_2+H_{\mathrm{sup}}+H_{\mathrm{int}}$ using the numerically exact diagonalization method. Our results for the in-gap Shiba-like states are presented in Fig.~\ref{fig:ShibaTwoSites}. Their presence is largely determined by the hopping amplitude $J_I$ even when the molecule acts as a magnet, i.e., for finite values of $\alpha\gamma$. It is logical as $J_I$ controls the penetration of spin-polarized electrons into the superconductor. By changing the value of $J_I$, the energy of the in-gap states can be modified and the crossing at zero energy can be observed. The behavior of energies in Fig.~\ref{fig:ShibaTwoSites}~(a) can be modified by changing the parameter $\alpha\gamma$, however, as long as $\alpha\gamma\neq 0$, only quantitative changes occur. Figure~\ref{fig:ShibaTwoSites}~(b) shows the probability amplitudes of the state at $J_I/J=2.0$, which demonstrate that the state is mainly located inside the superconductor.
All other essential features of Shiba states can be obtained by considering our model of a chiral molecule. A detailed matching to the experimental data requires a more elaborate study and will be presented elsewhere.

\textit{Discussion and conclusions:} We have shown that a 1D confined system in the presence of SOC and frictional dissipation can spatially separate particles of opposite spins~\footnote{It is worth noting that the spatial separation of spins implies a state, which appears more ordered than the state without dissipation, compare Figs.~\ref{fig:figure1}~a) and~\ref{fig:figure1}~b). According to the second law of thermodynamics, this implies that
if the electron is initially in a state shown in Fig.~\ref{fig:figure1}~b), then the entropy of the environment must increase during the dissipative process such that the change of the entropy of the total system (an electron plus environment) is non-negative.}. In our model, SOC and friction act in unison, so that pronounced spin polarization can be observed even if SOC is weak. We speculate that our findings may be related to the long-standing puzzle of the CISS effect, in which chiral molecules act as efficient spin filters despite weak SOC. The present model leads to spin polarization even in equilibrium, i.e., without currents, which sets it apart from the previously considered models of CISS. 

The friction term in our model is a phenomenological description of the electron-bath interactions. An important future milestone will be a derivation (from some microscopic Hamiltonian) of the proposed effective model that will establish a rigorous connection between friction and electron-bath interactions, as well as between molecular chirality and the direction of SOC.

As noted above, our findings are based on the assumption that the timescale of the spin coherence is large. This assumption is standard when considering propagation of electrons in a molecule, however, it should be scrutinized for a steady state considered here.  In future studies, we plan to investigate the steady state in the presence of spin-flip processes.

Our model suggests a metastable spin accumulation on the opposite ends of helical molecules. This accumulation should lead to a small localized magnetic moment, which can be probed experimentally. In solid-state systems, the magnetic moment may be measured using anomalous Hall measurements, as well as by tunneling spectroscopy of a Kondo resonance. We note that the spin accumulation at the edges of a chiral molecule could be highly relevant for molecule-molecule interactions in biological systems. Those interactions can be probed similarly to
interactions between spin-polarized chiral molecules and magnetic surfaces~\cite{Banerjee2018,Ziv2019}. 

Lastly, our model does not include temperature effects and further work is needed to understand the related finite-temperature physics. In particular, one could expect strong dependence of spin polarization on temperature. Indeed, spin coherence at low temperatures along with more rapid energy exchange between the bath and an electron at high temperatures might suggest the existence of some optimal temperature, which could be relevant for biological processes.

\begin{acknowledgments}
We thank Rafael Barfknecht for useful discussions. This work has received funding from the
European Union's Horizon 2020 research and innovation programme under the Marie Sk\l{}odowska-Curie Grant Agreement No. 754411 (A. G. and A. G. V.). M.~L. acknowledges support by the European Research Council (ERC) Starting Grant No. 801770 (ANGULON). Y.P. and O.M. acknowledge funding from the Nidersachsen Ministry of Science and Culture, and from the Academia Sinica Research Program. O.M. thanks support through the Harry de Jur Chair in Applied Science. 
\end{acknowledgments}

\bibliography{spinpolarizerbib}

\begin{thebibliography}{60}%
\makeatletter
\providecommand \@ifxundefined [1]{%
 \@ifx{#1\undefined}
}%
\providecommand \@ifnum [1]{%
 \ifnum #1\expandafter \@firstoftwo
 \else \expandafter \@secondoftwo
 \fi
}%
\providecommand \@ifx [1]{%
 \ifx #1\expandafter \@firstoftwo
 \else \expandafter \@secondoftwo
 \fi
}%
\providecommand \natexlab [1]{#1}%
\providecommand \enquote  [1]{``#1''}%
\providecommand \bibnamefont  [1]{#1}%
\providecommand \bibfnamefont [1]{#1}%
\providecommand \citenamefont [1]{#1}%
\providecommand \href@noop [0]{\@secondoftwo}%
\providecommand \href [0]{\begingroup \@sanitize@url \@href}%
\providecommand \@href[1]{\@@startlink{#1}\@@href}%
\providecommand \@@href[1]{\endgroup#1\@@endlink}%
\providecommand \@sanitize@url [0]{\catcode `\\12\catcode `\$12\catcode
  `\&12\catcode `\#12\catcode `\^12\catcode `\_12\catcode `\%12\relax}%
\providecommand \@@startlink[1]{}%
\providecommand \@@endlink[0]{}%
\providecommand \url  [0]{\begingroup\@sanitize@url \@url }%
\providecommand \@url [1]{\endgroup\@href {#1}{\urlprefix }}%
\providecommand \urlprefix  [0]{URL }%
\providecommand \Eprint [0]{\href }%
\providecommand \doibase [0]{http://dx.doi.org/}%
\providecommand \selectlanguage [0]{\@gobble}%
\providecommand \bibinfo  [0]{\@secondoftwo}%
\providecommand \bibfield  [0]{\@secondoftwo}%
\providecommand \translation [1]{[#1]}%
\providecommand \BibitemOpen [0]{}%
\providecommand \bibitemStop [0]{}%
\providecommand \bibitemNoStop [0]{.\EOS\space}%
\providecommand \EOS [0]{\spacefactor3000\relax}%
\providecommand \BibitemShut  [1]{\csname bibitem#1\endcsname}%
\let\auto@bib@innerbib\@empty
\bibitem [{\citenamefont {McGinley}\ and\ \citenamefont
  {Cooper}(2020)}]{McGinley2020}%
  \BibitemOpen
  \bibfield  {author} {\bibinfo {author} {\bibfnamefont {M.}~\bibnamefont
  {McGinley}}\ and\ \bibinfo {author} {\bibfnamefont {N.~R.}\ \bibnamefont
  {Cooper}},\ }\href {https://doi.org/10.1038/s41567-020-0956-z} {\bibfield
  {journal} {\bibinfo  {journal} {Nat. Phys.}\ }\textbf {\bibinfo {volume}
  {16}},\ \bibinfo {pages} {1181} (\bibinfo {year} {2020})}\BibitemShut
  {NoStop}%
\bibitem [{\citenamefont {{Lieu}}\ \emph {et~al.}(2021)\citenamefont {{Lieu}},
  \citenamefont {{McGinley}}, \citenamefont {{Shtanko}}, \citenamefont
  {{Cooper}},\ and\ \citenamefont {{Gorshkov}}}]{Lieu2021}%
  \BibitemOpen
  \bibfield  {author} {\bibinfo {author} {\bibfnamefont {S.}~\bibnamefont
  {{Lieu}}}, \bibinfo {author} {\bibfnamefont {M.}~\bibnamefont {{McGinley}}},
  \bibinfo {author} {\bibfnamefont {O.}~\bibnamefont {{Shtanko}}}, \bibinfo
  {author} {\bibfnamefont {N.~R.}\ \bibnamefont {{Cooper}}}, \ and\ \bibinfo
  {author} {\bibfnamefont {A.~V.}\ \bibnamefont {{Gorshkov}}},\ }\href@noop {}
  {\bibfield  {journal} {\bibinfo  {journal} {arXiv e-prints}\ ,\ \bibinfo
  {eid} {arXiv:2105.02888}} (\bibinfo {year} {2021})},\ \Eprint
  {http://arxiv.org/abs/2105.02888} {arXiv:2105.02888 [cond-mat.mes-hall]}
  \BibitemShut {NoStop}%
\bibitem [{\citenamefont {Yeganeh}\ \emph {et~al.}(2009)\citenamefont
  {Yeganeh}, \citenamefont {Ratner}, \citenamefont {Medina},\ and\
  \citenamefont {Mujica}}]{Yeganeh2009}%
  \BibitemOpen
  \bibfield  {author} {\bibinfo {author} {\bibfnamefont {S.}~\bibnamefont
  {Yeganeh}}, \bibinfo {author} {\bibfnamefont {M.~A.}\ \bibnamefont {Ratner}},
  \bibinfo {author} {\bibfnamefont {E.}~\bibnamefont {Medina}}, \ and\ \bibinfo
  {author} {\bibfnamefont {V.}~\bibnamefont {Mujica}},\ }\href {\doibase
  10.1063/1.3167404} {\bibfield  {journal} {\bibinfo  {journal} {J. Chem.
  Phys.}\ }\textbf {\bibinfo {volume} {131}},\ \bibinfo {pages} {014707}
  (\bibinfo {year} {2009})}\BibitemShut {NoStop}%
\bibitem [{\citenamefont {Medina}\ \emph {et~al.}(2012)\citenamefont {Medina},
  \citenamefont {L{\'o}pez}, \citenamefont {Ratner},\ and\ \citenamefont
  {Mujica}}]{Medina2012}%
  \BibitemOpen
  \bibfield  {author} {\bibinfo {author} {\bibfnamefont {E.}~\bibnamefont
  {Medina}}, \bibinfo {author} {\bibfnamefont {F.}~\bibnamefont {L{\'o}pez}},
  \bibinfo {author} {\bibfnamefont {M.~A.}\ \bibnamefont {Ratner}}, \ and\
  \bibinfo {author} {\bibfnamefont {V.}~\bibnamefont {Mujica}},\ }\href
  {\doibase 10.1209/0295-5075/99/17006} {\bibfield  {journal} {\bibinfo
  {journal} {EPL (Europhysics Letters)}\ }\textbf {\bibinfo {volume} {99}},\
  \bibinfo {pages} {17006} (\bibinfo {year} {2012})}\BibitemShut {NoStop}%
\bibitem [{\citenamefont {Varela}\ \emph {et~al.}(2014)\citenamefont {Varela},
  \citenamefont {Medina}, \citenamefont {Lopez},\ and\ \citenamefont
  {Mujica}}]{Varela2013}%
  \BibitemOpen
  \bibfield  {author} {\bibinfo {author} {\bibfnamefont {S.}~\bibnamefont
  {Varela}}, \bibinfo {author} {\bibfnamefont {E.}~\bibnamefont {Medina}},
  \bibinfo {author} {\bibfnamefont {F.}~\bibnamefont {Lopez}}, \ and\ \bibinfo
  {author} {\bibfnamefont {V.}~\bibnamefont {Mujica}},\ }\href {\doibase
  10.1088/0953-8984/26/1/015008} {\bibfield  {journal} {\bibinfo  {journal} {J.
  Phys.: Condens. Matter}\ }\textbf {\bibinfo {volume} {26}},\ \bibinfo {pages}
  {015008} (\bibinfo {year} {2014})}\BibitemShut {NoStop}%
\bibitem [{\citenamefont {Guo}\ and\ \citenamefont {Sun}(2012)}]{Guo2012}%
  \BibitemOpen
  \bibfield  {author} {\bibinfo {author} {\bibfnamefont {A.-M.}\ \bibnamefont
  {Guo}}\ and\ \bibinfo {author} {\bibfnamefont {Q.-f.}\ \bibnamefont {Sun}},\
  }\href {\doibase 10.1103/PhysRevLett.108.218102} {\bibfield  {journal}
  {\bibinfo  {journal} {Phys. Rev. Lett.}\ }\textbf {\bibinfo {volume} {108}},\
  \bibinfo {pages} {218102} (\bibinfo {year} {2012})}\BibitemShut {NoStop}%
\bibitem [{\citenamefont {Gutierrez}\ \emph {et~al.}(2012)\citenamefont
  {Gutierrez}, \citenamefont {D{\'\i}az}, \citenamefont {Naaman},\ and\
  \citenamefont {Cuniberti}}]{Gutierrez2012}%
  \BibitemOpen
  \bibfield  {author} {\bibinfo {author} {\bibfnamefont {R.}~\bibnamefont
  {Gutierrez}}, \bibinfo {author} {\bibfnamefont {E.}~\bibnamefont
  {D{\'\i}az}}, \bibinfo {author} {\bibfnamefont {R.}~\bibnamefont {Naaman}}, \
  and\ \bibinfo {author} {\bibfnamefont {G.}~\bibnamefont {Cuniberti}},\ }\href
  {\doibase 10.1103/PhysRevB.85.081404} {\bibfield  {journal} {\bibinfo
  {journal} {Phys. Rev. B}\ }\textbf {\bibinfo {volume} {85}},\ \bibinfo
  {pages} {081404} (\bibinfo {year} {2012})}\BibitemShut {NoStop}%
\bibitem [{\citenamefont {Gutierrez}\ \emph {et~al.}(2013)\citenamefont
  {Gutierrez}, \citenamefont {D{\'\i}az}, \citenamefont {Gaul}, \citenamefont
  {Brumme}, \citenamefont {Dom{\'\i}nguez-Adame},\ and\ \citenamefont
  {Cuniberti}}]{Gutierrez2013}%
  \BibitemOpen
  \bibfield  {author} {\bibinfo {author} {\bibfnamefont {R.}~\bibnamefont
  {Gutierrez}}, \bibinfo {author} {\bibfnamefont {E.}~\bibnamefont
  {D{\'\i}az}}, \bibinfo {author} {\bibfnamefont {C.}~\bibnamefont {Gaul}},
  \bibinfo {author} {\bibfnamefont {T.}~\bibnamefont {Brumme}}, \bibinfo
  {author} {\bibfnamefont {F.}~\bibnamefont {Dom{\'\i}nguez-Adame}}, \ and\
  \bibinfo {author} {\bibfnamefont {G.}~\bibnamefont {Cuniberti}},\ }\href
  {\doibase 10.1021/jp401705x} {\bibfield  {journal} {\bibinfo  {journal} {J.
  Phys. Chem. C}\ }\textbf {\bibinfo {volume} {117}},\ \bibinfo {pages} {22276}
  (\bibinfo {year} {2013})}\BibitemShut {NoStop}%
\bibitem [{\citenamefont {Guo}\ and\ \citenamefont {Sun}(2014)}]{Guo2014}%
  \BibitemOpen
  \bibfield  {author} {\bibinfo {author} {\bibfnamefont {A.-M.}\ \bibnamefont
  {Guo}}\ and\ \bibinfo {author} {\bibfnamefont {Q.-F.}\ \bibnamefont {Sun}},\
  }\href {\doibase 10.1103/PhysRevB.93.075407} {\bibfield  {journal} {\bibinfo
  {journal} {Proc. Natl. Acad. Sci. USA}\ }\textbf {\bibinfo {volume} {111}},\
  \bibinfo {pages} {11658} (\bibinfo {year} {2014})}\BibitemShut {NoStop}%
\bibitem [{\citenamefont {Ortix}(2015)}]{Ortix2015}%
  \BibitemOpen
  \bibfield  {author} {\bibinfo {author} {\bibfnamefont {C.}~\bibnamefont
  {Ortix}},\ }\href {\doibase 10.1103/PhysRevB.91.245412} {\bibfield  {journal}
  {\bibinfo  {journal} {Phys. Rev. B}\ }\textbf {\bibinfo {volume} {91}},\
  \bibinfo {pages} {245412} (\bibinfo {year} {2015})}\BibitemShut {NoStop}%
\bibitem [{\citenamefont {Matityahu}\ \emph {et~al.}(2016)\citenamefont
  {Matityahu}, \citenamefont {Utsumi}, \citenamefont {Aharony}, \citenamefont
  {Entin-Wohlman},\ and\ \citenamefont {Balseiro}}]{Matityahu2016}%
  \BibitemOpen
  \bibfield  {author} {\bibinfo {author} {\bibfnamefont {S.}~\bibnamefont
  {Matityahu}}, \bibinfo {author} {\bibfnamefont {Y.}~\bibnamefont {Utsumi}},
  \bibinfo {author} {\bibfnamefont {A.}~\bibnamefont {Aharony}}, \bibinfo
  {author} {\bibfnamefont {O.}~\bibnamefont {Entin-Wohlman}}, \ and\ \bibinfo
  {author} {\bibfnamefont {C.~A.}\ \bibnamefont {Balseiro}},\ }\href {\doibase
  10.1103/PhysRevB.93.075407} {\bibfield  {journal} {\bibinfo  {journal} {Phys.
  Rev. B}\ }\textbf {\bibinfo {volume} {93}},\ \bibinfo {pages} {075407}
  (\bibinfo {year} {2016})}\BibitemShut {NoStop}%
\bibitem [{\citenamefont {Michaeli}\ and\ \citenamefont
  {Naaman}(2019)}]{Michaeli2019}%
  \BibitemOpen
  \bibfield  {author} {\bibinfo {author} {\bibfnamefont {K.}~\bibnamefont
  {Michaeli}}\ and\ \bibinfo {author} {\bibfnamefont {R.}~\bibnamefont
  {Naaman}},\ }\href {\doibase 10.1021/acs.jpcc.9b05020} {\bibfield  {journal}
  {\bibinfo  {journal} {J. Phys. Chem. C}\ }\textbf {\bibinfo {volume} {123}},\
  \bibinfo {pages} {17043} (\bibinfo {year} {2019})}\BibitemShut {NoStop}%
\bibitem [{\citenamefont {Yang}\ \emph {et~al.}(2019)\citenamefont {Yang},
  \citenamefont {van~der Wal},\ and\ \citenamefont {van Wees}}]{Yang2019}%
  \BibitemOpen
  \bibfield  {author} {\bibinfo {author} {\bibfnamefont {X.}~\bibnamefont
  {Yang}}, \bibinfo {author} {\bibfnamefont {C.~H.}\ \bibnamefont {van~der
  Wal}}, \ and\ \bibinfo {author} {\bibfnamefont {B.~J.}\ \bibnamefont {van
  Wees}},\ }\href {\doibase 10.1103/PhysRevB.99.024418} {\bibfield  {journal}
  {\bibinfo  {journal} {Phys. Rev. B}\ }\textbf {\bibinfo {volume} {99}},\
  \bibinfo {pages} {024418} (\bibinfo {year} {2019})}\BibitemShut {NoStop}%
\bibitem [{\citenamefont {Geyer}\ \emph {et~al.}(2019)\citenamefont {Geyer},
  \citenamefont {Gutierrez}, \citenamefont {Mujica},\ and\ \citenamefont
  {Cuniberti}}]{Geyer2019}%
  \BibitemOpen
  \bibfield  {author} {\bibinfo {author} {\bibfnamefont {M.}~\bibnamefont
  {Geyer}}, \bibinfo {author} {\bibfnamefont {R.}~\bibnamefont {Gutierrez}},
  \bibinfo {author} {\bibfnamefont {V.}~\bibnamefont {Mujica}}, \ and\ \bibinfo
  {author} {\bibfnamefont {G.}~\bibnamefont {Cuniberti}},\ }\href {\doibase
  10.1021/acs.jpcc.9b07764} {\bibfield  {journal} {\bibinfo  {journal} {J.
  Phys. Chem. C}\ }\textbf {\bibinfo {volume} {123}},\ \bibinfo {pages} {27230}
  (\bibinfo {year} {2019})}\BibitemShut {NoStop}%
\bibitem [{\citenamefont {Gersten}\ \emph {et~al.}(2013)\citenamefont
  {Gersten}, \citenamefont {Kaasbjerg},\ and\ \citenamefont
  {Nitzan}}]{Gersten2013}%
  \BibitemOpen
  \bibfield  {author} {\bibinfo {author} {\bibfnamefont {J.}~\bibnamefont
  {Gersten}}, \bibinfo {author} {\bibfnamefont {K.}~\bibnamefont {Kaasbjerg}},
  \ and\ \bibinfo {author} {\bibfnamefont {A.}~\bibnamefont {Nitzan}},\ }\href
  {\doibase 10.1063/1.48209072} {\bibfield  {journal} {\bibinfo  {journal} {J.
  Chem. Phys.}\ }\textbf {\bibinfo {volume} {139}},\ \bibinfo {pages} {114111}
  (\bibinfo {year} {2013})}\BibitemShut {NoStop}%
\bibitem [{\citenamefont {Dalum}\ and\ \citenamefont
  {Hedeg{\aa}rd}(2019)}]{Dalum2019}%
  \BibitemOpen
  \bibfield  {author} {\bibinfo {author} {\bibfnamefont {S.}~\bibnamefont
  {Dalum}}\ and\ \bibinfo {author} {\bibfnamefont {P.}~\bibnamefont
  {Hedeg{\aa}rd}},\ }\href {\doibase 10.1021/acs.nanolett.9b01707} {\bibfield
  {journal} {\bibinfo  {journal} {Nano Lett.}\ }\textbf {\bibinfo {volume}
  {19}},\ \bibinfo {pages} {5253} (\bibinfo {year} {2019})}\BibitemShut
  {NoStop}%
\bibitem [{\citenamefont {Fransson}(2019)}]{Fransson2019}%
  \BibitemOpen
  \bibfield  {author} {\bibinfo {author} {\bibfnamefont {J.}~\bibnamefont
  {Fransson}},\ }\href {\doibase 10.1021/acs.jpclett.9b02929} {\bibfield
  {journal} {\bibinfo  {journal} {J. Phys. Chem. Lett.}\ }\textbf {\bibinfo
  {volume} {10}},\ \bibinfo {pages} {7126} (\bibinfo {year}
  {2019})}\BibitemShut {NoStop}%
\bibitem [{\citenamefont {Ghazaryan}\ \emph
  {et~al.}(2020{\natexlab{a}})\citenamefont {Ghazaryan}, \citenamefont
  {Paltiel},\ and\ \citenamefont {Lemeshko}}]{Ghazaryan2020}%
  \BibitemOpen
  \bibfield  {author} {\bibinfo {author} {\bibfnamefont {A.}~\bibnamefont
  {Ghazaryan}}, \bibinfo {author} {\bibfnamefont {Y.}~\bibnamefont {Paltiel}},
  \ and\ \bibinfo {author} {\bibfnamefont {M.}~\bibnamefont {Lemeshko}},\
  }\href {\doibase 10.1021/acs.jpcc.0c02584} {\bibfield  {journal} {\bibinfo
  {journal} {J. Phys. Chem. C}\ }\textbf {\bibinfo {volume} {124}},\ \bibinfo
  {pages} {11716} (\bibinfo {year} {2020}{\natexlab{a}})}\BibitemShut {NoStop}%
\bibitem [{\citenamefont {Du}\ \emph {et~al.}(2020)\citenamefont {Du},
  \citenamefont {Fu},\ and\ \citenamefont {Wu}}]{Du2020}%
  \BibitemOpen
  \bibfield  {author} {\bibinfo {author} {\bibfnamefont {G.-F.}\ \bibnamefont
  {Du}}, \bibinfo {author} {\bibfnamefont {H.-H.}\ \bibnamefont {Fu}}, \ and\
  \bibinfo {author} {\bibfnamefont {R.}~\bibnamefont {Wu}},\ }\href
  {https://doi.org/10.1103/PhysRevB.102.035431} {\bibfield  {journal} {\bibinfo
   {journal} {Phys. Rev. B}\ }\textbf {\bibinfo {volume} {102}},\ \bibinfo
  {pages} {035431} (\bibinfo {year} {2020})}\BibitemShut {NoStop}%
\bibitem [{\citenamefont {Ghazaryan}\ \emph
  {et~al.}(2020{\natexlab{b}})\citenamefont {Ghazaryan}, \citenamefont
  {Lemeshko},\ and\ \citenamefont {Volosniev}}]{Ghazaryan2020filtering}%
  \BibitemOpen
  \bibfield  {author} {\bibinfo {author} {\bibfnamefont {A.}~\bibnamefont
  {Ghazaryan}}, \bibinfo {author} {\bibfnamefont {M.}~\bibnamefont {Lemeshko}},
  \ and\ \bibinfo {author} {\bibfnamefont {A.~G.}\ \bibnamefont {Volosniev}},\
  }\href {https://doi.org/10.1038/s42005-020-00445-8} {\bibfield  {journal}
  {\bibinfo  {journal} {Communications Physics}\ }\textbf {\bibinfo {volume}
  {3}},\ \bibinfo {pages} {1} (\bibinfo {year}
  {2020}{\natexlab{b}})}\BibitemShut {NoStop}%
\bibitem [{\citenamefont {Utsumi}\ \emph {et~al.}(2020)\citenamefont {Utsumi},
  \citenamefont {Entin-Wohlman},\ and\ \citenamefont {Aharony}}]{Utsumi2020}%
  \BibitemOpen
  \bibfield  {author} {\bibinfo {author} {\bibfnamefont {Y.}~\bibnamefont
  {Utsumi}}, \bibinfo {author} {\bibfnamefont {O.}~\bibnamefont
  {Entin-Wohlman}}, \ and\ \bibinfo {author} {\bibfnamefont {A.}~\bibnamefont
  {Aharony}},\ }\href {\doibase 10.1103/PhysRevB.102.035445} {\bibfield
  {journal} {\bibinfo  {journal} {Phys. Rev. B}\ }\textbf {\bibinfo {volume}
  {102}},\ \bibinfo {pages} {035445} (\bibinfo {year} {2020})}\BibitemShut
  {NoStop}%
\bibitem [{\citenamefont {Fransson}(0)}]{Fransson2021}%
  \BibitemOpen
  \bibfield  {author} {\bibinfo {author} {\bibfnamefont {J.}~\bibnamefont
  {Fransson}},\ }\href {\doibase 10.1021/acs.nanolett.1c00183} {\bibfield
  {journal} {\bibinfo  {journal} {Nano Letters}\ }\textbf {\bibinfo {volume}
  {0}},\ \bibinfo {pages} {null} (\bibinfo {year} {0})},\ \bibinfo {note}
  {pMID: 33759530}\BibitemShut {NoStop}%
\bibitem [{\citenamefont {Liu}\ \emph {et~al.}(2021)\citenamefont {Liu},
  \citenamefont {Xiao}, \citenamefont {Koo},\ and\ \citenamefont
  {Yan}}]{Liu2021}%
  \BibitemOpen
  \bibfield  {author} {\bibinfo {author} {\bibfnamefont {Y.}~\bibnamefont
  {Liu}}, \bibinfo {author} {\bibfnamefont {J.}~\bibnamefont {Xiao}}, \bibinfo
  {author} {\bibfnamefont {J.}~\bibnamefont {Koo}}, \ and\ \bibinfo {author}
  {\bibfnamefont {B.}~\bibnamefont {Yan}},\ }\href
  {https://doi.org/10.1038/s41563-021-00924-5} {\bibfield  {journal} {\bibinfo
  {journal} {Nat. Mater}\ } (\bibinfo {year} {2021})}\BibitemShut {NoStop}%
\bibitem [{\citenamefont {G{\"o}hler}\ \emph {et~al.}(2011)\citenamefont
  {G{\"o}hler}, \citenamefont {Hamelbeck}, \citenamefont {Markus},
  \citenamefont {Kettner}, \citenamefont {Hanne}, \citenamefont {Vager},
  \citenamefont {Naaman},\ and\ \citenamefont {Zacharias}}]{Gohler2011}%
  \BibitemOpen
  \bibfield  {author} {\bibinfo {author} {\bibfnamefont {B.}~\bibnamefont
  {G{\"o}hler}}, \bibinfo {author} {\bibfnamefont {V.}~\bibnamefont
  {Hamelbeck}}, \bibinfo {author} {\bibfnamefont {T.}~\bibnamefont {Markus}},
  \bibinfo {author} {\bibfnamefont {M.}~\bibnamefont {Kettner}}, \bibinfo
  {author} {\bibfnamefont {G.}~\bibnamefont {Hanne}}, \bibinfo {author}
  {\bibfnamefont {Z.}~\bibnamefont {Vager}}, \bibinfo {author} {\bibfnamefont
  {R.}~\bibnamefont {Naaman}}, \ and\ \bibinfo {author} {\bibfnamefont
  {H.}~\bibnamefont {Zacharias}},\ }\href {\doibase 10.1126/science.1199339}
  {\bibfield  {journal} {\bibinfo  {journal} {Science}\ }\textbf {\bibinfo
  {volume} {331}},\ \bibinfo {pages} {894} (\bibinfo {year}
  {2011})}\BibitemShut {NoStop}%
\bibitem [{\citenamefont {Naaman}\ and\ \citenamefont
  {Waldeck}(2015)}]{Naaman2015}%
  \BibitemOpen
  \bibfield  {author} {\bibinfo {author} {\bibfnamefont {R.}~\bibnamefont
  {Naaman}}\ and\ \bibinfo {author} {\bibfnamefont {D.~H.}\ \bibnamefont
  {Waldeck}},\ }\href {\doibase 10.1146/annurev-physchem-040214-121554}
  {\bibfield  {journal} {\bibinfo  {journal} {Annu. Rev. Phys. Chem.}\ }\textbf
  {\bibinfo {volume} {66}},\ \bibinfo {pages} {263} (\bibinfo {year}
  {2015})}\BibitemShut {NoStop}%
\bibitem [{\citenamefont {Naaman}\ \emph {et~al.}(2019)\citenamefont {Naaman},
  \citenamefont {Paltiel},\ and\ \citenamefont {Waldeck}}]{naaman2019_review}%
  \BibitemOpen
  \bibfield  {author} {\bibinfo {author} {\bibfnamefont {R.}~\bibnamefont
  {Naaman}}, \bibinfo {author} {\bibfnamefont {Y.}~\bibnamefont {Paltiel}}, \
  and\ \bibinfo {author} {\bibfnamefont {D.~H.}\ \bibnamefont {Waldeck}},\
  }\href {https://doi.org/10.1038/s41570-019-0087-1} {\bibfield  {journal}
  {\bibinfo  {journal} {Nature Reviews Chemistry}\ }\textbf {\bibinfo {volume}
  {3}},\ \bibinfo {pages} {250} (\bibinfo {year} {2019})}\BibitemShut {NoStop}%
\bibitem [{\citenamefont {Kittel}(1953)}]{Kittel1953}%
  \BibitemOpen
  \bibfield  {author} {\bibinfo {author} {\bibfnamefont {C.}~\bibnamefont
  {Kittel}},\ }\href@noop {} {\emph {\bibinfo {title} {Introduction to Solid
  State Physics}}}\ (\bibinfo  {publisher} {University of Michigan: Wiley},\
  \bibinfo {year} {1953})\BibitemShut {NoStop}%
\bibitem [{\citenamefont {Senitzky}(1960)}]{Senitzky1960}%
  \BibitemOpen
  \bibfield  {author} {\bibinfo {author} {\bibfnamefont {I.~R.}\ \bibnamefont
  {Senitzky}},\ }\href {\doibase 10.1103/PhysRev.119.670} {\bibfield  {journal}
  {\bibinfo  {journal} {Phys. Rev.}\ }\textbf {\bibinfo {volume} {119}},\
  \bibinfo {pages} {670} (\bibinfo {year} {1960})}\BibitemShut {NoStop}%
\bibitem [{Note1()}]{Note1}%
  \BibitemOpen
  \bibinfo {note} {We focus on friction due to the resistance of the viscous
  medium. The non-contact friction~\cite {Persson2007} is not relevant for the
  present work.}\BibitemShut {Stop}%
\bibitem [{\citenamefont {Caldeira}\ and\ \citenamefont
  {Leggett}(1983)}]{Caldeira1983}%
  \BibitemOpen
  \bibfield  {author} {\bibinfo {author} {\bibfnamefont {A.}~\bibnamefont
  {Caldeira}}\ and\ \bibinfo {author} {\bibfnamefont {A.}~\bibnamefont
  {Leggett}},\ }\href {\doibase https://doi.org/10.1016/0378-4371(83)90013-4}
  {\bibfield  {journal} {\bibinfo  {journal} {Physica A: Statistical Mechanics
  and its Applications}\ }\textbf {\bibinfo {volume} {121}},\ \bibinfo {pages}
  {587 } (\bibinfo {year} {1983})}\BibitemShut {NoStop}%
\bibitem [{\citenamefont {Kanai}(1948)}]{Kanai1948}%
  \BibitemOpen
  \bibfield  {author} {\bibinfo {author} {\bibfnamefont {E.}~\bibnamefont
  {Kanai}},\ }\href {\doibase 10.1143/ptp/3.4.440} {\bibfield  {journal}
  {\bibinfo  {journal} {Progress of Theoretical Physics}\ }\textbf {\bibinfo
  {volume} {3}},\ \bibinfo {pages} {440} (\bibinfo {year} {1948})}\BibitemShut
  {NoStop}%
\bibitem [{\citenamefont {Kostin}(1972)}]{Kostin1972}%
  \BibitemOpen
  \bibfield  {author} {\bibinfo {author} {\bibfnamefont {M.~D.}\ \bibnamefont
  {Kostin}},\ }\href {\doibase 10.1063/1.1678812} {\bibfield  {journal}
  {\bibinfo  {journal} {The Journal of Chemical Physics}\ }\textbf {\bibinfo
  {volume} {57}},\ \bibinfo {pages} {3589} (\bibinfo {year}
  {1972})}\BibitemShut {NoStop}%
\bibitem [{\citenamefont {Hasse}(1975)}]{Hasse1975}%
  \BibitemOpen
  \bibfield  {author} {\bibinfo {author} {\bibfnamefont {R.~W.}\ \bibnamefont
  {Hasse}},\ }\href {\doibase 10.1063/1.522431} {\bibfield  {journal} {\bibinfo
   {journal} {Journal of Mathematical Physics}\ }\textbf {\bibinfo {volume}
  {16}},\ \bibinfo {pages} {2005} (\bibinfo {year} {1975})}\BibitemShut
  {NoStop}%
\bibitem [{\citenamefont {Albrecht}(1975)}]{Albrecht1975}%
  \BibitemOpen
  \bibfield  {author} {\bibinfo {author} {\bibfnamefont {K.}~\bibnamefont
  {Albrecht}},\ }\href {\doibase https://doi.org/10.1016/0370-2693(75)90283-X}
  {\bibfield  {journal} {\bibinfo  {journal} {Physics Letters B}\ }\textbf
  {\bibinfo {volume} {56}},\ \bibinfo {pages} {127 } (\bibinfo {year}
  {1975})}\BibitemShut {NoStop}%
\bibitem [{\citenamefont {Caldirola}\ and\ \citenamefont
  {Lugiato}(1982)}]{Caldirola1982}%
  \BibitemOpen
  \bibfield  {author} {\bibinfo {author} {\bibfnamefont {P.}~\bibnamefont
  {Caldirola}}\ and\ \bibinfo {author} {\bibfnamefont {L.}~\bibnamefont
  {Lugiato}},\ }\href {\doibase https://doi.org/10.1016/0378-4371(82)90242-4}
  {\bibfield  {journal} {\bibinfo  {journal} {Physica A: Statistical Mechanics
  and its Applications}\ }\textbf {\bibinfo {volume} {116}},\ \bibinfo {pages}
  {248 } (\bibinfo {year} {1982})}\BibitemShut {NoStop}%
\bibitem [{\citenamefont {Sun}\ and\ \citenamefont {Yu}(1995)}]{Sun1995}%
  \BibitemOpen
  \bibfield  {author} {\bibinfo {author} {\bibfnamefont {C.-P.}\ \bibnamefont
  {Sun}}\ and\ \bibinfo {author} {\bibfnamefont {L.-H.}\ \bibnamefont {Yu}},\
  }\href {\doibase 10.1103/PhysRevA.51.1845} {\bibfield  {journal} {\bibinfo
  {journal} {Phys. Rev. A}\ }\textbf {\bibinfo {volume} {51}},\ \bibinfo
  {pages} {1845} (\bibinfo {year} {1995})}\BibitemShut {NoStop}%
\bibitem [{\citenamefont {Schuch}(1999)}]{Schuch1999}%
  \BibitemOpen
  \bibfield  {author} {\bibinfo {author} {\bibfnamefont {D.}~\bibnamefont
  {Schuch}},\ }\href {\doibase
  https://doi.org/10.1002/(SICI)1097-461X(1999)72:6<537::AID-QUA1>3.0.CO;2-Q}
  {\bibfield  {journal} {\bibinfo  {journal} {International Journal of Quantum
  Chemistry}\ }\textbf {\bibinfo {volume} {72}},\ \bibinfo {pages} {537}
  (\bibinfo {year} {1999})}\BibitemShut {NoStop}%
\bibitem [{\citenamefont {Alpern}\ \emph {et~al.}(2016)\citenamefont {Alpern},
  \citenamefont {Katzir}, \citenamefont {Yochelis}, \citenamefont {Katz},
  \citenamefont {Paltiel},\ and\ \citenamefont {Millo}}]{Alpern2016}%
  \BibitemOpen
  \bibfield  {author} {\bibinfo {author} {\bibfnamefont {H.}~\bibnamefont
  {Alpern}}, \bibinfo {author} {\bibfnamefont {E.}~\bibnamefont {Katzir}},
  \bibinfo {author} {\bibfnamefont {S.}~\bibnamefont {Yochelis}}, \bibinfo
  {author} {\bibfnamefont {N.}~\bibnamefont {Katz}}, \bibinfo {author}
  {\bibfnamefont {Y.}~\bibnamefont {Paltiel}}, \ and\ \bibinfo {author}
  {\bibfnamefont {O.}~\bibnamefont {Millo}},\ }\href {\doibase
  10.1088/1367-2630/18/11/113048} {\bibfield  {journal} {\bibinfo  {journal}
  {New J. Phys.}\ }\textbf {\bibinfo {volume} {18}},\ \bibinfo {pages} {113048}
  (\bibinfo {year} {2016})}\BibitemShut {NoStop}%
\bibitem [{\citenamefont {Shapira}\ \emph {et~al.}(2018)\citenamefont
  {Shapira}, \citenamefont {Alpern}, \citenamefont {Yochelis}, \citenamefont
  {Lee}, \citenamefont {Kaun}, \citenamefont {Paltiel}, \citenamefont {Koren},\
  and\ \citenamefont {Millo}}]{Shapira2018}%
  \BibitemOpen
  \bibfield  {author} {\bibinfo {author} {\bibfnamefont {T.}~\bibnamefont
  {Shapira}}, \bibinfo {author} {\bibfnamefont {H.}~\bibnamefont {Alpern}},
  \bibinfo {author} {\bibfnamefont {S.}~\bibnamefont {Yochelis}}, \bibinfo
  {author} {\bibfnamefont {T.-K.}\ \bibnamefont {Lee}}, \bibinfo {author}
  {\bibfnamefont {C.-C.}\ \bibnamefont {Kaun}}, \bibinfo {author}
  {\bibfnamefont {Y.}~\bibnamefont {Paltiel}}, \bibinfo {author} {\bibfnamefont
  {G.}~\bibnamefont {Koren}}, \ and\ \bibinfo {author} {\bibfnamefont
  {O.}~\bibnamefont {Millo}},\ }\href
  {https://doi.org/10.1103/PhysRevB.98.214513} {\bibfield  {journal} {\bibinfo
  {journal} {Phys. Rev. B}\ }\textbf {\bibinfo {volume} {98}},\ \bibinfo
  {pages} {214513} (\bibinfo {year} {2018})}\BibitemShut {NoStop}%
\bibitem [{\citenamefont {Alpern}\ \emph {et~al.}(2019)\citenamefont {Alpern},
  \citenamefont {Yavilberg}, \citenamefont {Dvir}, \citenamefont {Sukenik},
  \citenamefont {Klang}, \citenamefont {Yochelis}, \citenamefont {Cohen},
  \citenamefont {Grosfeld}, \citenamefont {Steinberg}, \citenamefont {Paltiel}
  \emph {et~al.}}]{Alpern2019}%
  \BibitemOpen
  \bibfield  {author} {\bibinfo {author} {\bibfnamefont {H.}~\bibnamefont
  {Alpern}}, \bibinfo {author} {\bibfnamefont {K.}~\bibnamefont {Yavilberg}},
  \bibinfo {author} {\bibfnamefont {T.}~\bibnamefont {Dvir}}, \bibinfo {author}
  {\bibfnamefont {N.}~\bibnamefont {Sukenik}}, \bibinfo {author} {\bibfnamefont
  {M.}~\bibnamefont {Klang}}, \bibinfo {author} {\bibfnamefont
  {S.}~\bibnamefont {Yochelis}}, \bibinfo {author} {\bibfnamefont
  {H.}~\bibnamefont {Cohen}}, \bibinfo {author} {\bibfnamefont
  {E.}~\bibnamefont {Grosfeld}}, \bibinfo {author} {\bibfnamefont
  {H.}~\bibnamefont {Steinberg}}, \bibinfo {author} {\bibfnamefont
  {Y.}~\bibnamefont {Paltiel}},  \emph {et~al.},\ }\href
  {https://doi.org/10.1021/acs.nanolett.9b01552} {\bibfield  {journal}
  {\bibinfo  {journal} {Nano Lett.}\ }\textbf {\bibinfo {volume} {19}},\
  \bibinfo {pages} {5167} (\bibinfo {year} {2019})}\BibitemShut {NoStop}%
\bibitem [{\citenamefont {Ben~Dor}\ \emph {et~al.}(2017)\citenamefont
  {Ben~Dor}, \citenamefont {Yochelis}, \citenamefont {Radko}, \citenamefont
  {Vankayala}, \citenamefont {Capua}, \citenamefont {Capua}, \citenamefont
  {Yang}, \citenamefont {Baczewski}, \citenamefont {Parkin}, \citenamefont
  {Naaman},\ and\ \citenamefont {Paltiel}}]{BenDor2017}%
  \BibitemOpen
  \bibfield  {author} {\bibinfo {author} {\bibfnamefont {O.}~\bibnamefont
  {Ben~Dor}}, \bibinfo {author} {\bibfnamefont {S.}~\bibnamefont {Yochelis}},
  \bibinfo {author} {\bibfnamefont {A.}~\bibnamefont {Radko}}, \bibinfo
  {author} {\bibfnamefont {K.}~\bibnamefont {Vankayala}}, \bibinfo {author}
  {\bibfnamefont {E.}~\bibnamefont {Capua}}, \bibinfo {author} {\bibfnamefont
  {A.}~\bibnamefont {Capua}}, \bibinfo {author} {\bibfnamefont {S.-H.}\
  \bibnamefont {Yang}}, \bibinfo {author} {\bibfnamefont {L.~T.}\ \bibnamefont
  {Baczewski}}, \bibinfo {author} {\bibfnamefont {S.~S.~P.}\ \bibnamefont
  {Parkin}}, \bibinfo {author} {\bibfnamefont {R.}~\bibnamefont {Naaman}}, \
  and\ \bibinfo {author} {\bibfnamefont {Y.}~\bibnamefont {Paltiel}},\ }\href
  {\doibase 10.1038/ncomms14567} {\bibfield  {journal} {\bibinfo  {journal}
  {Nature Communications}\ }\textbf {\bibinfo {volume} {8}},\ \bibinfo {pages}
  {14567} (\bibinfo {year} {2017})}\BibitemShut {NoStop}%
\bibitem [{\citenamefont {Sukenik}\ \emph {et~al.}(2020)\citenamefont
  {Sukenik}, \citenamefont {Tassinari}, \citenamefont {Yochelis}, \citenamefont
  {Millo}, \citenamefont {Baczewski},\ and\ \citenamefont
  {Paltiel}}]{Sukenik2020}%
  \BibitemOpen
  \bibfield  {author} {\bibinfo {author} {\bibfnamefont {N.}~\bibnamefont
  {Sukenik}}, \bibinfo {author} {\bibfnamefont {F.}~\bibnamefont {Tassinari}},
  \bibinfo {author} {\bibfnamefont {S.}~\bibnamefont {Yochelis}}, \bibinfo
  {author} {\bibfnamefont {O.}~\bibnamefont {Millo}}, \bibinfo {author}
  {\bibfnamefont {L.~T.}\ \bibnamefont {Baczewski}}, \ and\ \bibinfo {author}
  {\bibfnamefont {Y.}~\bibnamefont {Paltiel}},\ }\href@noop {} {\bibfield
  {journal} {\bibinfo  {journal} {Molecules}\ }\textbf {\bibinfo {volume}
  {25}},\ \bibinfo {pages} {24} (\bibinfo {year} {2020})}\BibitemShut {NoStop}%
\bibitem [{\citenamefont {Rashba}(2003)}]{Rashba2003}%
  \BibitemOpen
  \bibfield  {author} {\bibinfo {author} {\bibfnamefont {E.~I.}\ \bibnamefont
  {Rashba}},\ }\href {\doibase 10.1103/PhysRevB.68.241315} {\bibfield
  {journal} {\bibinfo  {journal} {Phys. Rev. B}\ }\textbf {\bibinfo {volume}
  {68}},\ \bibinfo {pages} {241315} (\bibinfo {year} {2003})}\BibitemShut
  {NoStop}%
\bibitem [{\citenamefont {Fajardo}\ \emph {et~al.}(2017)\citenamefont
  {Fajardo}, \citenamefont {Z\"ulicke},\ and\ \citenamefont
  {Winkler}}]{Fajardo2017}%
  \BibitemOpen
  \bibfield  {author} {\bibinfo {author} {\bibfnamefont {E.~A.}\ \bibnamefont
  {Fajardo}}, \bibinfo {author} {\bibfnamefont {U.}~\bibnamefont {Z\"ulicke}},
  \ and\ \bibinfo {author} {\bibfnamefont {R.}~\bibnamefont {Winkler}},\ }\href
  {\doibase 10.1103/PhysRevB.96.155304} {\bibfield  {journal} {\bibinfo
  {journal} {Phys. Rev. B}\ }\textbf {\bibinfo {volume} {96}},\ \bibinfo
  {pages} {155304} (\bibinfo {year} {2017})}\BibitemShut {NoStop}%
\bibitem [{Note2()}]{Note2}%
  \BibitemOpen
  \bibinfo {note} {Note that the average value of $p$ is zero for $\alpha =0$.
  In this case, there is no frictional force, i.e., $W=0$, and the steady-state
  solutions of $H$ are simply the eigenstates of a square well. Therefore,
  non-vanishing spin-orbit coupling is essential for the main results of
  present work.}\BibitemShut {Stop}%
\bibitem [{Note3()}]{Note3}%
  \BibitemOpen
  \bibinfo {note} {If the potential $V$ is not mirror symmetric, then there is
  no double degeneracy associated with spin. This occurs due to dissipation,
  which explicitly breaks time-reversal system, and brings the system outside
  the region of applicability of the Kramers degeneracy theorem.}\BibitemShut
  {Stop}%
\bibitem [{Note4()}]{Note4}%
  \BibitemOpen
  \bibinfo {note} {See Ref.~\cite {Hasse1975} for a detailed comparison of
  different phenomenological models of friction for systems without
  SOC.}\BibitemShut {Stop}%
\bibitem [{\citenamefont {Moroz}\ and\ \citenamefont
  {Barnes}(1999)}]{Moroz1999}%
  \BibitemOpen
  \bibfield  {author} {\bibinfo {author} {\bibfnamefont {A.~V.}\ \bibnamefont
  {Moroz}}\ and\ \bibinfo {author} {\bibfnamefont {C.~H.~W.}\ \bibnamefont
  {Barnes}},\ }\href {\doibase 10.1103/PhysRevB.60.14272} {\bibfield  {journal}
  {\bibinfo  {journal} {Phys. Rev. B}\ }\textbf {\bibinfo {volume} {60}},\
  \bibinfo {pages} {14272} (\bibinfo {year} {1999})}\BibitemShut {NoStop}%
\bibitem [{\citenamefont {Kumar}\ \emph {et~al.}(2017)\citenamefont {Kumar},
  \citenamefont {Capua}, \citenamefont {Kesharwani}, \citenamefont {Martin},
  \citenamefont {Sitbon}, \citenamefont {Waldeck},\ and\ \citenamefont
  {Naaman}}]{Kumar2017}%
  \BibitemOpen
  \bibfield  {author} {\bibinfo {author} {\bibfnamefont {A.}~\bibnamefont
  {Kumar}}, \bibinfo {author} {\bibfnamefont {E.}~\bibnamefont {Capua}},
  \bibinfo {author} {\bibfnamefont {M.~K.}\ \bibnamefont {Kesharwani}},
  \bibinfo {author} {\bibfnamefont {J.~M.}\ \bibnamefont {Martin}}, \bibinfo
  {author} {\bibfnamefont {E.}~\bibnamefont {Sitbon}}, \bibinfo {author}
  {\bibfnamefont {D.~H.}\ \bibnamefont {Waldeck}}, \ and\ \bibinfo {author}
  {\bibfnamefont {R.}~\bibnamefont {Naaman}},\ }\href {\doibase
  10.1073/pnas.1611467114} {\bibfield  {journal} {\bibinfo  {journal} {Proc.
  Natl. Acad. Sci. USA}\ }\textbf {\bibinfo {volume} {114}},\ \bibinfo {pages}
  {2474} (\bibinfo {year} {2017})}\BibitemShut {NoStop}%
\bibitem [{\citenamefont {Kettner}\ \emph {et~al.}(2018)\citenamefont
  {Kettner}, \citenamefont {Maslyuk}, \citenamefont {N{\"u}renberg},
  \citenamefont {Seibel}, \citenamefont {Gutierrez}, \citenamefont {Cuniberti},
  \citenamefont {Ernst},\ and\ \citenamefont {Zacharias}}]{Kettner2018}%
  \BibitemOpen
  \bibfield  {author} {\bibinfo {author} {\bibfnamefont {M.}~\bibnamefont
  {Kettner}}, \bibinfo {author} {\bibfnamefont {V.~V.}\ \bibnamefont
  {Maslyuk}}, \bibinfo {author} {\bibfnamefont {D.}~\bibnamefont
  {N{\"u}renberg}}, \bibinfo {author} {\bibfnamefont {J.}~\bibnamefont
  {Seibel}}, \bibinfo {author} {\bibfnamefont {R.}~\bibnamefont {Gutierrez}},
  \bibinfo {author} {\bibfnamefont {G.}~\bibnamefont {Cuniberti}}, \bibinfo
  {author} {\bibfnamefont {K.-H.}\ \bibnamefont {Ernst}}, \ and\ \bibinfo
  {author} {\bibfnamefont {H.}~\bibnamefont {Zacharias}},\ }\href {\doibase
  10.1021/acs.jpclett.8b00208} {\bibfield  {journal} {\bibinfo  {journal} {J.
  Phys. Chem. Lett.}\ }\textbf {\bibinfo {volume} {9}},\ \bibinfo {pages}
  {2025} (\bibinfo {year} {2018})}\BibitemShut {NoStop}%
\bibitem [{\citenamefont {Nitzan}(2006)}]{Nitzan2006Book}%
  \BibitemOpen
  \bibfield  {author} {\bibinfo {author} {\bibfnamefont {A.}~\bibnamefont
  {Nitzan}},\ }\href@noop {} {\emph {\bibinfo {title} {Chemical dynamics in
  condensed phases: relaxation, transfer and reactions in condensed molecular
  systems}}}\ (\bibinfo  {publisher} {Oxford university press},\ \bibinfo
  {year} {2006})\BibitemShut {NoStop}%
\bibitem [{\citenamefont {Zhou}\ \emph {et~al.}(2010)\citenamefont {Zhou},
  \citenamefont {Lei}, \citenamefont {Wang}, \citenamefont {Pei}, \citenamefont
  {Cao},\ and\ \citenamefont {Wang}}]{Zhou2010}%
  \BibitemOpen
  \bibfield  {author} {\bibinfo {author} {\bibfnamefont {Y.}~\bibnamefont
  {Zhou}}, \bibinfo {author} {\bibfnamefont {T.}~\bibnamefont {Lei}}, \bibinfo
  {author} {\bibfnamefont {L.}~\bibnamefont {Wang}}, \bibinfo {author}
  {\bibfnamefont {J.}~\bibnamefont {Pei}}, \bibinfo {author} {\bibfnamefont
  {Y.}~\bibnamefont {Cao}}, \ and\ \bibinfo {author} {\bibfnamefont
  {J.}~\bibnamefont {Wang}},\ }\href {https://doi.org/10.1002/adma.200904171}
  {\bibfield  {journal} {\bibinfo  {journal} {Advanced Materials}\ }\textbf
  {\bibinfo {volume} {22}},\ \bibinfo {pages} {1484} (\bibinfo {year}
  {2010})}\BibitemShut {NoStop}%
\bibitem [{\citenamefont {Shen}\ and\ \citenamefont {Chen}(2012)}]{Shen2012}%
  \BibitemOpen
  \bibfield  {author} {\bibinfo {author} {\bibfnamefont {Y.}~\bibnamefont
  {Shen}}\ and\ \bibinfo {author} {\bibfnamefont {C.-F.}\ \bibnamefont
  {Chen}},\ }\href {https://doi.org/10.1021/cr200087r} {\bibfield  {journal}
  {\bibinfo  {journal} {Chemical reviews}\ }\textbf {\bibinfo {volume} {112}},\
  \bibinfo {pages} {1463} (\bibinfo {year} {2012})}\BibitemShut {NoStop}%
\bibitem [{\citenamefont {Xu}\ \emph {et~al.}(2019)\citenamefont {Xu},
  \citenamefont {Li}, \citenamefont {Ricciarelli}, \citenamefont {Wang},
  \citenamefont {Mosconi}, \citenamefont {Yuan}, \citenamefont {De~Angelis},
  \citenamefont {Zakeeruddin}, \citenamefont {Gr{\"a}tzel},\ and\ \citenamefont
  {Wang}}]{Xu2019}%
  \BibitemOpen
  \bibfield  {author} {\bibinfo {author} {\bibfnamefont {N.}~\bibnamefont
  {Xu}}, \bibinfo {author} {\bibfnamefont {Y.}~\bibnamefont {Li}}, \bibinfo
  {author} {\bibfnamefont {D.}~\bibnamefont {Ricciarelli}}, \bibinfo {author}
  {\bibfnamefont {J.}~\bibnamefont {Wang}}, \bibinfo {author} {\bibfnamefont
  {E.}~\bibnamefont {Mosconi}}, \bibinfo {author} {\bibfnamefont
  {Y.}~\bibnamefont {Yuan}}, \bibinfo {author} {\bibfnamefont {F.}~\bibnamefont
  {De~Angelis}}, \bibinfo {author} {\bibfnamefont {S.~M.}\ \bibnamefont
  {Zakeeruddin}}, \bibinfo {author} {\bibfnamefont {M.}~\bibnamefont
  {Gr{\"a}tzel}}, \ and\ \bibinfo {author} {\bibfnamefont {P.}~\bibnamefont
  {Wang}},\ }\href {https://doi.org/10.1016/j.isci.2019.04.031} {\bibfield
  {journal} {\bibinfo  {journal} {Iscience}\ }\textbf {\bibinfo {volume}
  {15}},\ \bibinfo {pages} {234} (\bibinfo {year} {2019})}\BibitemShut
  {NoStop}%
\bibitem [{\citenamefont {Shiba}(1968)}]{Shiba1968}%
  \BibitemOpen
  \bibfield  {author} {\bibinfo {author} {\bibfnamefont {H.}~\bibnamefont
  {Shiba}},\ }\href {\doibase 10.1143/PTP.40.435} {\bibfield  {journal}
  {\bibinfo  {journal} {Progress of Theoretical Physics}\ }\textbf {\bibinfo
  {volume} {40}},\ \bibinfo {pages} {435} (\bibinfo {year} {1968})}\BibitemShut
  {NoStop}%
\bibitem [{\citenamefont {Zhu}(2016)}]{Zhu2016book}%
  \BibitemOpen
  \bibfield  {author} {\bibinfo {author} {\bibfnamefont {J.-X.}\ \bibnamefont
  {Zhu}},\ }\href {https://link.springer.com/book/10.1007/978-3-319-31314-6}
  {\emph {\bibinfo {title} {Bogoliubov-de Gennes Method and Its
  Applications}}}\ (\bibinfo  {publisher} {Springer},\ \bibinfo {year}
  {2016})\BibitemShut {NoStop}%
\bibitem [{Note5()}]{Note5}%
  \BibitemOpen
  \bibinfo {note} {It is worth noting that the spatial separation of spins
  implies a state, which appears more ordered than the state without
  dissipation, compare Figs.~\ref {fig:figure1}~a) and~\ref {fig:figure1}~b).
  According to the second law of thermodynamics, this implies that if the
  electron is initially in a state shown in Fig.~\ref {fig:figure1}~b), then
  the entropy of the environment must increase during the dissipative process
  such that the change of the entropy of the total system (an electron plus
  environment) is non-negative.}\BibitemShut {Stop}%
\bibitem [{\citenamefont {Banerjee-Ghosh}\ \emph {et~al.}(2018)\citenamefont
  {Banerjee-Ghosh}, \citenamefont {Dor}, \citenamefont {Tassinari},
  \citenamefont {Capua}, \citenamefont {Yochelis}, \citenamefont {Capua},
  \citenamefont {Yang}, \citenamefont {Parkin}, \citenamefont {Sarkar},
  \citenamefont {Kronik}, \citenamefont {Baczewski}, \citenamefont {Naaman},\
  and\ \citenamefont {Paltiel}}]{Banerjee2018}%
  \BibitemOpen
  \bibfield  {author} {\bibinfo {author} {\bibfnamefont {K.}~\bibnamefont
  {Banerjee-Ghosh}}, \bibinfo {author} {\bibfnamefont {O.~B.}\ \bibnamefont
  {Dor}}, \bibinfo {author} {\bibfnamefont {F.}~\bibnamefont {Tassinari}},
  \bibinfo {author} {\bibfnamefont {E.}~\bibnamefont {Capua}}, \bibinfo
  {author} {\bibfnamefont {S.}~\bibnamefont {Yochelis}}, \bibinfo {author}
  {\bibfnamefont {A.}~\bibnamefont {Capua}}, \bibinfo {author} {\bibfnamefont
  {S.-H.}\ \bibnamefont {Yang}}, \bibinfo {author} {\bibfnamefont {S.~S.}\
  \bibnamefont {Parkin}}, \bibinfo {author} {\bibfnamefont {S.}~\bibnamefont
  {Sarkar}}, \bibinfo {author} {\bibfnamefont {L.}~\bibnamefont {Kronik}},
  \bibinfo {author} {\bibfnamefont {L.~T.}\ \bibnamefont {Baczewski}}, \bibinfo
  {author} {\bibfnamefont {R.}~\bibnamefont {Naaman}}, \ and\ \bibinfo {author}
  {\bibfnamefont {Y.}~\bibnamefont {Paltiel}},\ }\href {\doibase
  10.1126/science.aar4265} {\bibfield  {journal} {\bibinfo  {journal}
  {Science}\ }\textbf {\bibinfo {volume} {360}},\ \bibinfo {pages} {1331}
  (\bibinfo {year} {2018})}\BibitemShut {NoStop}%
\bibitem [{\citenamefont {Ziv}\ \emph {et~al.}(2019)\citenamefont {Ziv},
  \citenamefont {Saha}, \citenamefont {Alpern}, \citenamefont {Sukenik},
  \citenamefont {Baczewski}, \citenamefont {Yochelis}, \citenamefont {Reches},\
  and\ \citenamefont {Paltiel}}]{Ziv2019}%
  \BibitemOpen
  \bibfield  {author} {\bibinfo {author} {\bibfnamefont {A.}~\bibnamefont
  {Ziv}}, \bibinfo {author} {\bibfnamefont {A.}~\bibnamefont {Saha}}, \bibinfo
  {author} {\bibfnamefont {H.}~\bibnamefont {Alpern}}, \bibinfo {author}
  {\bibfnamefont {N.}~\bibnamefont {Sukenik}}, \bibinfo {author} {\bibfnamefont
  {L.~T.}\ \bibnamefont {Baczewski}}, \bibinfo {author} {\bibfnamefont
  {S.}~\bibnamefont {Yochelis}}, \bibinfo {author} {\bibfnamefont
  {M.}~\bibnamefont {Reches}}, \ and\ \bibinfo {author} {\bibfnamefont
  {Y.}~\bibnamefont {Paltiel}},\ }\href {\doibase
  https://doi.org/10.1002/adma.201904206} {\bibfield  {journal} {\bibinfo
  {journal} {Advanced Materials}\ }\textbf {\bibinfo {volume} {31}},\ \bibinfo
  {pages} {1904206} (\bibinfo {year} {2019})}\BibitemShut {NoStop}%
\bibitem [{\citenamefont {Volokitin}\ and\ \citenamefont
  {Persson}(2007)}]{Persson2007}%
  \BibitemOpen
  \bibfield  {author} {\bibinfo {author} {\bibfnamefont {A.~I.}\ \bibnamefont
  {Volokitin}}\ and\ \bibinfo {author} {\bibfnamefont {B.~N.~J.}\ \bibnamefont
  {Persson}},\ }\href {\doibase 10.1103/RevModPhys.79.1291} {\bibfield
  {journal} {\bibinfo  {journal} {Rev. Mod. Phys.}\ }\textbf {\bibinfo {volume}
  {79}},\ \bibinfo {pages} {1291} (\bibinfo {year} {2007})}\BibitemShut
  {NoStop}%
\end{thebibliography}%
\bibliographystyle{apsrev4-1}

\end{document}